\documentclass[sigconf]{acmart}

\AtBeginDocument{%
  \providecommand\BibTeX{{%
    \normalfont B\kern-0.5em{\scshape i\kern-0.25em b}\kern-0.8em\TeX}}}
    
\copyrightyear{2023}
\acmYear{2023}
\setcopyright{acmlicensed}\acmConference[CHI '23]{Proceedings of the 2023 CHI Conference on Human Factors in Computing Systems}{April 23--28, 2023}{Hamburg, Germany}
\acmBooktitle{2023 CHI Conference on Human Factors in Computing Systems (CHI '23), April 28, 2023, Hamburg, Germany}
\acmPrice{15.00}
\acmDOI{10.1145/3544548.3581354}
\acmISBN{978-1-4503-9421-5/23/04}

\usepackage{booktabs}
\usepackage{graphicx}
\usepackage{caption}
\usepackage{lipsum}

\begin{document}

\title[Stakeholder-Centered AI Design]{Stakeholder-Centered AI Design: Co-Designing Worker Tools with Gig Workers through Data Probes}

\author{Angie Zhang}
\affiliation{
  \institution{School of Information \\ The University of Texas at Austin}
   \city{Austin}
   \state{TX}
  \country{United States}}
  \email{angie.zhang@austin.utexas.edu}

\author{Alexander Boltz}
\authornote{The second author conducted this work as a research associate at the University of Texas at Austin's School of Information.}
\affiliation{
  \institution{Human Centered Design \& Eng. \\ University of Washington}
   \city{Seattle}
   \state{WA}
  \country{United States}
}
\email{aboltz@uw.edu}

\author{Jonathan Lynn}
\affiliation{
  \institution{School of Information \\ The University of Texas at Austin}
   \city{Austin}
   \state{TX}
  \country{United States}
}
\email{jonathan.lynn@utexas.edu}

\author{Chun-Wei Wang}
\affiliation{
  \institution{School of Information \\ The University of Texas at Austin}
   \city{Austin}
   \state{TX}
  \country{United States}
}
\email{waynewang@utexas.edu}

\author{Min Kyung Lee}
\affiliation{
  \institution{School of Information \\ The University of Texas at Austin}
   \city{Austin}
   \state{TX}
  \country{United States}
}
\email{minkyung.lee@austin.utexas.edu}

\renewcommand{\shortauthors}{Zhang et al.}

\newcommand{\anedit}[1]{{\color{black}#1}}
\newcommand{\aledit}[1]{{\color{black}#1}}
\newcommand\jedit[1]{{\color{black} #1}} 

\begin{abstract}
AI technologies continue to advance from digital assistants to assisted decision-making. However, designing AI remains a challenge given its unknown outcomes and uses. One way to expand AI design is by centering stakeholders in the design process. We conduct co-design sessions with gig workers to explore the design of gig worker-centered tools as informed by their driving patterns, decisions, and personal contexts. Using workers' own data as well as city-level data, we create probes---interactive data visuals---that participants explore to surface the well-being and positionalities that shape their work strategies. \anedit{We describe participant insights and corresponding AI design considerations surfaced from data probes about: 1) workers’ well-being trade-offs and positionality constraints, 2) factors that impact well-being beyond those in the data probes, and 3) instances of unfair algorithmic management. We discuss the implications for designing data probes and using them to elevate worker-centered AI design as well as for worker advocacy.}

\end{abstract}

\ccsdesc[500]{Human-centered computing~Human computer interaction (HCI)}
\keywords{Gig Work, AI Design, Co-Design, Data Probes, Worker Well-Being, Worker-Centered Worker Tools}


\maketitle

\section{Introduction}



The advancement of AI has enabled a spectrum of innovative products and functions. From self-driving vehicles to mental health chatbots \cite{lee2020designing, bae2021social}, monitoring bank fraud to determining medical diagnoses \cite{kermany2018identifying}---when designed well, AI can be an assistive tool for humans in navigating their tasks. 

In the workplace, AI has increasingly automated repetitive, manual tasks and improved worker productivity and efficiency. Here, AI often manifests as algorithmic management---algorithms that take on managerial functions to oversee, assign tasks, and evaluate workers \cite{lee2015working, jarrahi2021algorithmic}. For example, workplace tools such as Microsoft Suites or productivity apps include monitoring features so employers can track worker efficiency and throughput \cite{sandler2020microsoft, ajunwa2017limitless}. Some workplaces may even continue to track and collect data on employees through these applications when they are off the clock \cite{ajunwa2018algorithms}. These practices of algorithmic management are especially prevalent in gig work where data about workers is collected in droves and overbearing algorithms dictate worker tasks \cite{rosenblat2016algorithmic}. 

Though vast amounts of data are collected through these applications on behalf of companies without consideration of the users, exploring one’s own data can be incredibly informative and empowering. Personal informatics (PI) research around how individuals desire to track, reflect on, and draw insights from their own data has shown how this can assist people in understanding their patterns \cite{rooksby2014personal} and even initiate behavior change \cite{walsh2014stepcity}. Additionally, exploring and reflecting on their real data can support users in imagining future uses of their data, surfacing ideas that designers or practitioners might not otherwise come up with on their own \cite{holstein2020replay}. 

With the ubiquity of digital data, PI researchers have explored how individuals review and reflect on their own data collected by social media platforms \cite{gulotta2017digital, lustig2022designing}. Researchers often use design probes, objects intended to help engage users in exploring open-ended questions \cite{wallace2013making}. Though often physical objects \cite{gaver1999design}, design probes may also take form as digital artifacts and even incorporate data as a way to encourage users in investigating questions and design. For example, by presenting data to users through design probes, \citet{gulotta2017digital} demonstrate how probes can help individuals both surface insights about patterns in their data and inform designers on how to design digital systems. Similarly, \citet{subramonyam2021towards} demonstrated that these design probes using data, which they call "data probes", can also solidify AI system boundaries: they show how designers and engineers surfaced concrete use cases and limitations when considering user data while designing AI. These examples show promising direction toward the use of data for not only supporting individuals in understanding their own data but also advancing AI design. However, apart from \cite{gulotta2017digital}, most existing work has explored how data probes can assist stakeholders in understanding their own data or practitioners in designing AI as opposed to how they might support stakeholders in co-designing AI. 

We explore this question---how to design AI with stakeholders to center their well-being and personal contexts---in the context of gig workers. Gig work platforms, where 16\% of Americans were estimated to work in 2021 \cite{anderson2021state}, is a fitting domain for us to explore our question due to the intense datafication of gig workers. With their work managed through mobile applications, a litany of data is collected about workers, from their hours worked and earnings, to more granular information such as their location even when they are not working. This presents a unique opportunity to explore how to co-design AI tools with workers by helping them harness their own work data through the exploration of data probes. We define data probes as digital objects which use historical data to help participants recall past or current patterns and events as they complete co-design activities.

We design five data probes using workers' own data---four visualizations and one AI work planning prototype---that workers interact with to surface their work patterns, well-being concerns, and positionality. During individual co-design sessions with 12 rideshare drivers, drivers reflect on well-being and positionality before interacting with the data probes created from their own work data as well as city-level aggregate data. For each probe, we introduce drivers to the metrics presented in the visualization and ask questions to encourage their exploration. We discover drivers use data probes as boundary objects to describe their work patterns and contexts, suggesting implications for AI design rooted in these contexts. Using probes, drivers 1) shared the well-being trade-offs and positionality factors they face in their work, 2) identified factors beyond those represented in their data probes which impact their work and well-being, and 3) identified instances of algorithmic management in their data. For each finding, we also share corresponding AI design considerations such as the creation of assistive AI to track and reflect on a driver’s well-being goals. Finally, we discuss the implications of our study on the design of data probes and the use of data probes for elevating worker expertise in AI design, including revealing individual contexts to inform AI design, as well as the potential for worker advocacy.

\section{Related Work} 
We first describe the challenges researchers have faced when designing AI or ML and the approaches that have been taken thus far to expand our understanding of designing AI with the use of data. We then describe approaches advancing PI and stakeholder-centered AI in the context of gig work to motivate our design and methodology.

\subsection{Designing AI: Challenges and Recent Approaches}
\label{Designing AI}
Despite the increasing prevalence of AI technologies, the challenge of designing AI persists \cite{dove2017ux,holmquist2017intelligence,yang2018machine,yang2020re}. Because of its dynamic nature, AI often resists the standard prototyping methods designers are used to, requiring necessary adaptations \cite{yang2019sketching}, and often leading to costly prototyping \cite{dove2017ux,yang2018machine,yang2020re}. For example, \citet{yang2019sketching} explored how designers used sketching to design NLP systems, revealing how sketching approaches had to be modified to support NLP designers. Additionally, designers often face uncertainty around AI capabilities \cite{yang2020re, dove2017ux}, potentially due to a gap in AI knowledge, which contributes to their struggles in envisioning AI use cases \cite{dove2017ux, yang2018machine}. Though this gap can often be bridged by designers working with developers to quickly understand the feasibility of their ideas \cite{yang2018machine, dove2017ux, yildirim2022experienced, ozenc2010support}, designers may not always have access to these individuals while working with AI.

One domain of research has investigated how the use of data may be able to assist with designing AI or digital systems. Researchers have employed data-based prototyping methods. For example, \citet{holstein2020replay} employed historical data through "Replay Enactments" to conduct feature prototyping, simulating experiences for teachers on technical systems. \citet{zhang2023deliberating} also used historical data with focus groups to elicit stakeholder feedback about future organizational decision-making practices. \citet{subramonyam2021towards} used data probes where designer-engineer teams considered end-user data to surface use cases and outcomes of AI. Others have studied how to use social media data with users to explore the design of digital experiences \cite{lustig2022designing, gulotta2017digital}. Though they focused on designing a physical product rather than AI, \citet{bogers2016connected} used data as a design material to improve physical prototypes. Through interviews and a diary study, they describe how the collection and analysis of qualitative data allowed them to deepen their understanding of quantitative data (i.e., sensor-collected data) which was largely meaningless on its own. This method of data-enabled design allows for greater immersion of data probes, resulting in more engaging designs. These methods have primarily employed data for simulating experiences, technologies, and behaviors. We identify an opportunity to expand on the idea of data probes put forth by \citet{subramonyam2021towards} to explore how stakeholders' own data may be explored and interpreted by them in order to design AI.

\subsection{Personal Informatics and the Gig Economy} \label{personal informatics}

As our study seeks to create data probes that support stakeholders in reflecting on their data patterns, we turn to personal informatics \aledit{(PI)} literature, a well-explored field in a variety of domains, particularly \anedit{for its support of} individual user reflection. In their influential model, \citet{li2010stage} proposed using 5 stages to design PI systems. Significant research has sought to assist individuals \anedit{in the latter two stages---reflection and action---}when bridging the gap between users’ goals and data \cite{munson2012mindfulness, cordeiro2015barriers, caraban201923, cho2022reflection}: these papers emphasize the importance of care in messaging taken between 'positive' and 'negative' nudges, and ongoing reflective support to create the ability for equally nuanced and personal reflections for users. \citet{you2021go} explored these two stages with drivers and their partners, finding that the combination of technology-sensing probes (wearables) with social sense-making (having drivers and their partners share reflections about the drivers' behaviors) was promising for encouraging work-life balance for drivers. They facilitated drivers in reflection through summary tables of tracked data and diary entries, suggesting future work for increasing the granularity of data analysis for drivers.

However, attention must also be paid to creating understandable and actionable measures for participants who are inexperienced with data, otherwise, researchers run the risk of participants abandoning their learned practices as soon as the study concludes \cite{boulard2018analysis, epstein2016beyond}. \citet{boulard2018analysis} specifically suggested \textit{contextual} and \textit{understandable} quantifications for users, perhaps even at the cost of simplicity. For that reason, we chose to not only create \anedit{data probes} based on individual participants' own data, but also present them side-by-side with local \anedit{city-level} data, so that they would be able to place their behaviors in the context of their peers. We predict that this will be an effective method of not only contextualization, but also as a self-reflective exercise, comparing self to others. 

\aledit{Algorithmic management is a particularly difficult, yet important, domain to examine PI data.} \anedit{Compared to traditional PI apps, the goals of rideshare apps for data collection center around maximizing profit rather than helping users, such as drivers, understand their data patterns; thus platform data practices are intentionally opaque. This leaves drivers to} \aledit{collectively sensemake gamified metrics \cite{vasudevan2022gamification, davalos_bennett_2022}.} Although there are some PI tools for rideshare drivers---including apps such as Gridwise, Hurdlr, or Stride \cite{ets-hokin_2022}---these apps have several challenges. These include paywalls or 'premium' features \anedit{for personalized insights}, different features between apps, mixed functionality across apps and platforms, \anedit{complexity of using} separate applications \anedit{simultaneously}, importing data, and \anedit{drivers} interpreting features and statistics on their own. As a result, drivers often have to calculate statistics themselves mentally or use simpler data analysis tools, limiting their ability to learn more advanced insights \cite{zhang2022algorithmic}. 

\anedit{Drawing on the goals of these tools and recent data privacy policy changes, we identify the potential to democratize personalized data tools for gig workers and for researchers to study the effect of access to these tools on well-being and performance. For example, the EU's General Data Protection Regulation (GDPR) ensures the user's right to their own data including on gig work platforms. This opens up the possibility for gig workers to investigate and harness their own data for exploration and insights. While the United States may not have policies on the same scale, a number of companies have shifted towards open data practices by allowing users to download their own data (e.g., Facebook, Google, Uber). This presents a prime opportunity for researchers to investigate the design of AI in the domain of digital work.}

\subsection{Stakeholder Involvement in AI Design} 
Past research on human-centered design methods for AI has primarily focused on assisting UX or ML practitioners, for example, \citet{yang2018investigating} interviewing UX practitioners for their challenges and practices when using AI as design material, \citet{subramonyam2021towards} conducting co-design sessions to see how data probes can support designer-engineer teams developing AI, and \citet{yildirim2022experienced} exploring the practices of enterprise designers working on cross-functional teams. In recent years though, calls have been made to use human-centered AI design methods to address the ethical issues that arise when AI is created void of stakeholder contexts to inform potential outcomes and harms of a technology \cite{wolf2018changing, loi2019co}. Resources and guidelines have been published to support designers in the pursuit of human-centered AI that is more inclusive of impacted stakeholders \cite{aragon2022human, shneiderman2022human, d2020data, costanza2020design, xu2019toward}.

Researchers have conducted interviews or workshops \cite{stapleton2022imagining, brown2019toward} and co-design or participatory design sessions \cite{zhang2022algorithmic, park2022designing} with impacted stakeholders and shown how this can generate AI design ideas based on stakeholders' own experiences and situations. Researchers have also worked specifically with gig workers. \citet{zhang2022algorithmic} held co-design sessions with rideshare drivers to surface drivers' ideas for how to address or redesign algorithmic management; \aledit{\citet{bates2021lessons} used co-design with courier drivers to examine their complex relationship with data, finding that reliance on PI tracking tools like Strava repurposed for gig work, can help fill the pieces left missing by the platform; and \citet{alvarez2022design} utilized design fiction as a means of co-designing systems to support freelancers.} \aledit{Co-design has also been used as a method for creating PI systems \cite{potapov2020lifemosaic, yao2019defending, kildea2019design}, however, most such studies focus on users as consumers. Worker PI data, encompassing their job and livelihood, is a meaningful direction for our work.}

\anedit{Integrating these methods to design AI for gig work is particularly relevant given the lack of user context in algorithmic management or worker-related models.} \aledit{Uber and Lyft currently use an advanced dynamic matching and pricing algorithm to connect drivers with riders, often in just seconds \cite{yan2020dynamic}. While highly efficient, these algorithms also require advanced decision-making from drivers, usually before they've even completed their previous task. Named \textit{dynamic waiting}, this has negative impacts on driver well-being 
by including very limited driver input to indicate work preferences \cite{zhang2022algorithmic}.} \anedit{Separate from platforms, researchers such as \cite{calacci2022bargaining} and \cite{pan2019dissecting} have developed algorithms using worker data to understand their earnings or how to improve them. However, these models do not currently incorporate worker context, which can lead to tools workers cannot use or miss the opportunity to discover unique worker-centered questions.} 

These instances and others centering stakeholders \cite{holstein2019co, lin2021engaging, alvarez2022design, ayobi2021machine, lee2019webuildai, zhu2018value} illustrate the potential in working with stakeholders to inform, through personal contexts and experiences, how to design AI. 

\section{Background: Partnership with Independent Drivers Guild} 
In the Summer of 2021, we began working with the co-founder of \aledit{Independent Drivers Guild of Illinois} in the U.S. to better understand the well-being challenges and algorithmic management faced by drivers for our then-ongoing study design. Shortly after, we brought him on in a compensated position as a community advisor to periodically seek his feedback about whether our research direction aligned with what would be most useful for drivers. In Fall 2021, following our first study, we presented the results to our community advisor to get his opinion about our ideas for follow-up research directions as well as discuss his ideas given his direct driver advocacy work. He was particularly interested in the data-driven ideas of drivers from our study, suggesting the potential for using data to assist drivers in comparing their individual metrics with city-level drivers. For example, he conceptualized an AI tool that could provide a weekly plan to the driver, customized \aledit{to} an individual's driving habits for where and when they should work given their goals and constraints. He introduced us to Chicago's public dataset of rideshare data and encouraged us to explore ways this could be used to assist drivers. 

This was a key factor in guiding our research direction toward how to design AI with gig workers through \anedit{data probes} created from individual data and city-level data. We began by using public city data to create data visuals on Tableau about rideshare earnings. We then reached out to our community advisor in Winter 2021 to gain feedback around these visuals and his thoughts on which ones drivers may find useful to explore, and ultimately moved forward with the visual that he felt was the most beneficial---a preliminary \anedit{work planning} tool to predict weekly driving earnings. We also began exploring \aledit{drivers' individual} data in order to see how this data could be used to help drivers surface ideas from their work to design AI. We learned the Uber platform allows drivers to request their data, and we reached out to two previous study participants in Spring 2022 who graciously agreed to provide their driving data in a compensated pilot exploring individual data.\footnote{These pilot participants are excluded from Table \ref{participanttable}}%

From these pilots, we gained a better sense of how the drivers might benefit from exploring their data using data probes---most prominently, we noted that both drivers naturally used the visualizations they saw (see Appendix for Figures \ref{pilot1} and \ref{pilot2}) to reflect and narrate specific events they recalled (e.g., a trip that went to Milwaukee), temporal patterns they had internalized (e.g., bar closing times), and well-being or personal situations that affected their driving patterns (e.g., a family member passing away from COVID). Additionally, one of the drivers used the data to explain specific ways he wanted to review his data to investigate whether certain patterns he followed led to successful earnings or not. For example, he explained he anticipated a drop in summer ridership \aledit{volume} and devised a tactic to try to increase his tips and subsequently his earnings per trip. He explained that reviewing his data for this time period could help him determine whether that idea had a tangible effect: "It'd be nice to know if, if I'm---if that extra effort is having an impact."

These interactions inspired us to revise the data visualizations to their current version in our study, where we focus on \anedit{using data probes} to support drivers' reflection and storytelling and surface specific contexts of their driving well-being and strategy formulation to inform AI design. In section \ref{DesignDataProbe}, we describe our design goals, principles, and challenges that we address in formulating our session activities and data probes \anedit{in order to reveal} sensitizing concepts for practitioners designing AI.

\section{Designing Data Probes} 
\label{DesignDataProbe}
\begin{table*}[h]
\begin{tabular}{@{}clcl@{}}
\toprule
\multicolumn{1}{c}{\textbf{No.}} & \multicolumn{1}{c}{\textbf{Data Probe Name}} & \multicolumn{1}{c}{\textbf{Data Type}} & \multicolumn{1}{l}{\textbf{Description}}  \\ \midrule

1 & \begin{tabular}[c]{@{}l@{}}Driving Animation \\ \textit{(Animation)}\end{tabular} & Personal & \begin{tabular}[c]{@{}l@{}}Visceral animated gif showing a specific day\\ of a worker’s movement patterns on a map of their city\end{tabular} \\ \midrule
2 & \begin{tabular}[c]{@{}l@{}}Neighborhood Map \\\textit{(Map)}\end{tabular} & \begin{tabular}[c]{@{}c@{}}Personal\\ \& \\ City-level\end{tabular}  & \begin{tabular}[c]{@{}l@{}}Interactive map showing the neighborhoods of where \\pick-ups have occurred. Drivers can hover or click on\\ neighborhoods to see statistics such as when and how \\many trips occurred, average fare, and miles per trip.\end{tabular} \\ \midrule
3 & Calendar & \begin{tabular}[c]{@{}c@{}}Personal \\ \& \\ City-level\end{tabular} & \begin{tabular}[c]{@{}l@{}}Intuitive monthly and weekly calendars that display \\earning trends and breakdown of trips for a \\specific day or day of the week.\end{tabular} \\ \midrule
4 & Hourly & \begin{tabular}[c]{@{}c@{}}Personal\\ \& \\ City-level\end{tabular}  & \begin{tabular}[c]{@{}l@{}}Interactive bar charts that use personal or city-level \\ data to display trends around earnings and trips \\ depending on hour of the day.\end{tabular} \\ \midrule
5 & \begin{tabular}[c]{@{}l@{}}Work Planner \\\textit{(Planner)}\end{tabular} & City-level & \begin{tabular}[c]{@{}l@{}}Interactive prototype that allows a user to input work\\ parameters (e.g., hours, days, and neighborhoods\\ worked) and uses city-level data to provide an \\ estimation of base earnings, tips, and mileage for the \\ driver in a week\end{tabular} \\ \bottomrule

\end{tabular}
\vspace{2mm}
\captionsetup{width=0.75\textwidth}
\caption{Summary of the 5 Data Probes. Further details about the data used to create them can be found in Sections \ref{DesignDataProbe} and \ref{DataExplanation}.}
\label{table:dataprobetable}

\end{table*}

We were drawn to how detailed and reflective our pilot participants’ insights were about their data, from the specific rides they recalled to how their personal contexts shaped their patterns. We wondered how the contexts drivers surface from exploring their data might inform AI design or AI-based tools for drivers. \anedit{Designing AI can refer to different design phases---such as design space exploration which \citet{zhang2022algorithmic} pursued through scenario-based co-design sessions with drivers to explore algorithmic management interventions---or it may refer to surfacing specific components for AI tools such as data types, feature engineering, prediction and optimization goals, and constraints. One set of ideas drivers surfaced in \cite{zhang2022algorithmic} centered around using their own data to discover insights and support their well-being. Building off of this, our pilot studies,} and past work around AI design with data probes (Section \ref{Designing AI}), our goal became to co-design AI---explore components of AI tools such as feature engineering and constraints---with drivers through data probes---visualizations created using their own data. To design effective data probes that drivers feel empowered to use, regardless of experience in data analysis, we identified design principles to adhere to.

\subsection{Supporting Reflection and Action}
For our first principle, we wanted to ensure the data probes supported drivers' ease of understanding when reflecting on their data and identifying subsequent actions they can take \cite{li2010stage}. Cognizant of how data analysis can be daunting to participants without analysis experience \cite{moore2021exploring}, we wanted to design data probes to reflect drivers’ data in forms that are familiar to them, ideally, formats that can remind them of everyday representations.

This led us to design one data probe that displays driver data \anedit{on a calendar}. We chose this based on how drivers in our previous study often discussed reviewing earnings at the end of the day or week. Additionally, given the ubiquity of calendars, we hoped a calendar \anedit{data probe} could support drivers in reflecting on their personal calendar of activities and \anedit{discuss} how life events or contexts influenced their driving.

Based on discussions with our community advisor and the fact that maps are the most common artifact drivers interact with, we created two \anedit{more} data probes: 1) an interactive map of Chicago with earnings trends by neighborhood, and 2) an animation displaying a day of the drivers’ trips. \anedit{Drivers we spoke to often referenced the maps within rideshare apps when describing the information they use for decision-making. However, these maps only display immediate demand of certain regions rather than insights about a driver's personal patterns. Thus, we created an interactive map to support drivers in generating location-specific insights. Similarly, we presented drivers with \anedit{personalized} animations that trace their point-by-point movements to make their data more tangible for reflections.} 

We included the work planning tool, Planner, \anedit{as a data probe} as our community advisor believed it could be beneficial for drivers \anedit{and because it presents a use case that can help drivers anchor their ideas for AI components such as predictions or constraints to consider}. \anedit{Drivers can input driving parameters such as times and places they work, and weekly expenses to obtain weekly predictions} (e.g., total earnings, trips, and miles in a week). \anedit{This is displayed as} a table and text summary to support individual preferences of reading and reflecting on data. \anedit{See Figure \ref{fig:scheduler-full} for the Planner's full input and output options and Section \ref{scheduler-description} for how the Planner works.}

For our final data probe representing drivers' earnings based on trip pick-up hour, we cycled through many iterations, finally landing on a bar chart due to its simplicity as a visual compared to the other options. 

To further support drivers in reflection, we incorporated heat maps into the calendar and map \anedit{data probes}. The colors allow participants to easily identify profitable days or neighborhoods. We also designed questions to begin task-based and then become open-ended, as opposed to asking open-ended questions like "what are your thoughts?" which can be intimidating given the data they are presented with.
\anedit{These data probes are summarized in Table \ref{table:dataprobetable} as well as Figure \ref{fig:overviewfig} which situates each probe in the co-design session. We describe the data populating these probes in Section \ref{DataExplanation}.}

\subsection{Centering Context}
Our second principle centers on individual contexts that drivers hold which influence their work preferences, patterns, and limitations. Having the drivers discuss their strategies and patterns within the contexts that bind their work may make AI capabilities more certain and outputs more attuned to what drivers need. Our approach is largely informed by suggestions from \citet{d2020data}. We use visualizations \anedit{for our data probes} to make patterns visible, followed by probing questions about the data to uncover the context behind the presented patterns.

In order to elevate driver context, we incorporate well-being and positionality considerations prior to exploring data probes and then throughout questions as they explore data probes. To encourage drivers to consider the personal contexts that influence their work, we introduced participants to the concepts of well-being and positionality at the beginning of each session. For this study, we focus on three types of well-being: physical, psychological, and financial, building from \citet{hickok2022framework}'s definitions of the terms \cite{lee2021participatory}. We also have drivers complete a positionality activity---"the personal values, views, and location in time and space that influence how one engages with and understands the world" \cite{kearney_2022}---by having them consider what factors advantage and disadvantage them as a driver. Some of these factors are derived from existing positionality wheels (e.g., "Level of Education, Age, Race \& Ethnicity") \cite{noel2021learning} while others we added specific to gig work ("Own vs. Rent My Car", "Single vs. Multi-Platform Worker"). Later, as drivers interact with data probes, we ask how their work patterns affect their well-being and remind them of the positionality factors they identified to see if/how any of their factors \anedit{impact their choice to} work or not at a certain time or place.

\section{Methods}
\subsection{Study Participants}
\begin{table*}[h]
\resizebox{\textwidth}{!}{%
\begin{tabular}{@{}lccccccccc@{}}
\toprule
\bf P &
  \bf Age &
  \bf Race &
  \bf Gender &
  \bf Education &
  \begin{tabular}[c]{@{}c@{}}\bf Gig-work\\ \bf Tenure\end{tabular} &
  \begin{tabular}[c]{@{}c@{}}\bf Hours\\ \bf per Week\end{tabular} &
  \begin{tabular}[c]{@{}c@{}}\bf Household\\ \bf Size\end{tabular} &
  \begin{tabular}[c]{@{}c@{}}\bf Other\\ \bf Job?\end{tabular} &
  \begin{tabular}[c]{@{}c@{}}\bf Platforms Worked\\ \bf (primary first)\end{tabular} \\ \midrule
1 &
  33 &
  Asian &
  Male &
  Prefer not to answer &
  5 years+ &
  30 - 45 &
  \begin{tabular}[c]{@{}c@{}}Prefer not\\ to answer\end{tabular} &
  No &
  \begin{tabular}[c]{@{}c@{}}Uber, Lyft Amazon\\ Flex, DoorDash, \\ Grubhub\end{tabular} \\ \midrule
2  & 32 & Hispanic & Male   & Some graduate school      & 2-5 years   & 45 - 60 & 2 & No  & Uber, Lyft, DoorDash \\ \midrule
3  & 35 & Asian    & Male   & Postgraduate/prof. degree & 6-12 months & 45 - 60 & 3 & No  & Uber                 \\ \midrule
4  & 56 & White    & Male   & Some college, no degree   & 5 years+    & 0 - 15  & 2 & Yes & Uber                 \\ \midrule
5  & 50 & Asian    & Male   & Some college, no degree   & 5 years+    & 30 - 45 & 4 & No  & Uber                 \\ \midrule
6  & 55 & White    & Male   & Some college, no degree   & 5 years+    & 30 - 45 & 1 & No  & Uber, Lyft           \\ \midrule
7  & 43 & White    & Male   & Postgraduate/prof. degree & 2-5 years   & 30 - 45 & 1 & No  & Uber, Lyft           \\ \midrule
8  & 49 & White    & Female & Postgraduate/prof. degree & 2-5 years   & 15 - 30 & 1 & No  & Lyft, Uber           \\ \midrule
9  & 35 & Asian    & Male   & High school               & 5 Years+    & 45 - 60 & 3 & No  & Uber                 \\ \midrule
10 & 40 & White    & Male   & Postgraduate/prof. degree & 5 Years+    & 0 - 15  & 3 & Yes & Uber, Lyft           \\ \midrule
11 & 54 & White    & Female & Two-year associate degree & 5 Years+    & 30 - 45 & 3 & Yes & Lyft, Instacart      \\ \midrule
12 & 45 & White    & Male   & Some college, no degree   & 2-5 years   & 45 - 60 & 7 & No  & Uber, Lyft, Rodeo    \\ \bottomrule
\end{tabular}%
}
\vspace{0.2cm}
\caption{Participant Demographics and Work History} \label{participanttable}
\end{table*}
\subsubsection{Participant Criteria}
We recruited active Chicago Uber drivers. We required participants to be Uber drivers primarily because Uber is the only rideshare platform we identified as allowing drivers to easily request and download their data. The participants had to be actively driving because one of the \anedit{data probes} we showed was based on their geodata which Uber only provides for the past 30 days. Finally, the participants had to be working in Chicago because of \anedit{available} city-level rideshare data with earnings. 

We used multiple recruitment methods: Facebook Ads, posts in Facebook Groups, posts on subreddits, Reddit ads, word-of-mouth, and emails through our partner channels. 
We distributed a pre-screening survey. In addition to basic demographic data, we asked respondents to self-classify the permanence of their gig work careers \cite{dunn2020making}, whether they had another job, and how much they relied upon their rideshare income ("nice to have, but not essential to my budget" or "essential for meeting basic needs" \cite{smith2016gig}). Our research team reached out to respondents, detailing instructions on how to download their Uber data and share with the research team.

\subsubsection{Participants}
We had $n$=12 total participants. Their average age was 43.9 years (SD = 2.57). 7 of our 12 (58.3\%) participants identified as White, 4 (33.3\%) as Asian, and 1 as Hispanic (8.3\%). 10 of our participants (83.3\%) identified as male, with the remaining 2 identifying as female (16.7\%). 11 of our participants (91.7\%) have been working for rideshare platforms for 2 years or more. 10 participants (83.3\%) also reported working at least 30 hours a week on gig platforms, with the same 10 reporting that their gig work income was "essential for meeting basic needs"; 2 participants reported that their income was "nice to have, but not essential to my budget".

\subsection{Data Probe Creation} 
\label{DataExplanation}
\anedit{We first describe the data sources we used and the four visualization data probes. We then explain the fifth data probe, the Planner---its mechanics and computations. Most data probes were created primarily using Tableau, an interactive data visualization platform.}

\subsubsection{Data Sources to Create Data Probes}
\hfill

\begin{figure}[!ht]
\captionsetup{width=0.7\textwidth}
  \includegraphics[width=.9\linewidth]{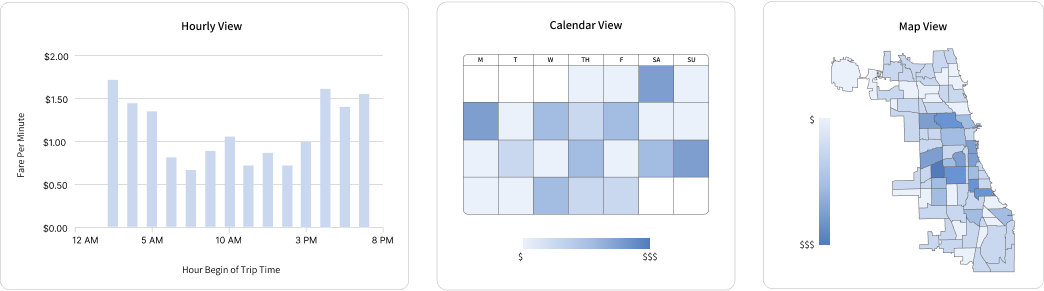}
  \caption{Individual Data Probes}
  \label{fig:personalfig}
\end{figure}

\textbf{Individual Data.} To create individual \anedit{data probes} (see Figure \ref{fig:personalfig}) we utilized the driver's own Uber data\footnote{Uber's methodology can be found here: https://help.uber.com/driving-and-delivering/article/request-your-personal-uber-data?nodeId=fbf08e68-65ba-456b-9bc6-1369eb9d2c44}. 

\textbf{Personal Trip Animation Data Probe.} \anedit{We created a personalized trip animation for each driver}, combining trips with location data from the past 30 days. With this file, we generated a time-lapse animation of participants' movements overlaid on a map of Chicago using Unfolded.ai\footnote{https://www.unfolded.ai/ is a geospatial analytics platform}. For the session, we isolated the animation to one recent day of trips. The purpose of the animation was to give participants a visceral representation of their movement data as they described their general work strategy. 

\textbf{Individual Hourly, Calendar, and Map Data Probes.} Next, participants saw their hourly \anedit{data probe}, a bar chart displaying the average fare per minute at each hour of the day for the month of June. Then, they reviewed their calendar displaying similar metrics, summarized for each day since June 2022. For the map \anedit{data probe}, participants saw two personalized maps, colored in varying shades based on the average fares per minute in each neighborhood. The first showed their common pick-up locations (as well as total pick-ups per area upon hover). The second showed their common drop-off locations. These maps were interactive: selecting a pick-up neighborhood altered the drop-off map to display only drop-off locations for trips beginning in the selected neighborhood.

\textbf{City-Level Data.} The data for city-level \anedit{data probes} came directly from the City of Chicago\footnote{https://data.cityofchicago.org/Transportation/Transportation-Network-Providers-Trips/m6dm-c72p}. Available back to 2018, we limited our dataset to the most recently available month (June 2022) because of size and processing power limitations. To prepare the dataset for analysis, we used Pandas---a Python-based data manipulation tool---to classify all the pick-up and drop-off coordinates into Chicago's official neighborhoods, as defined by the City of Chicago\footnote{https://data.cityofchicago.org/Facilities-Geographic-Boundaries/Boundaries-Neighborhoods/bbvz-uum9}. Next, we used Pandas to add time-appropriate weather data acquired from VisualCrossing, a live weather API\footnote{https://www.visualcrossing.com/weather-history/chicago/us}. 

\begin{figure}[!ht]
\captionsetup{width=0.7\textwidth}
  \includegraphics[width=.9\linewidth]{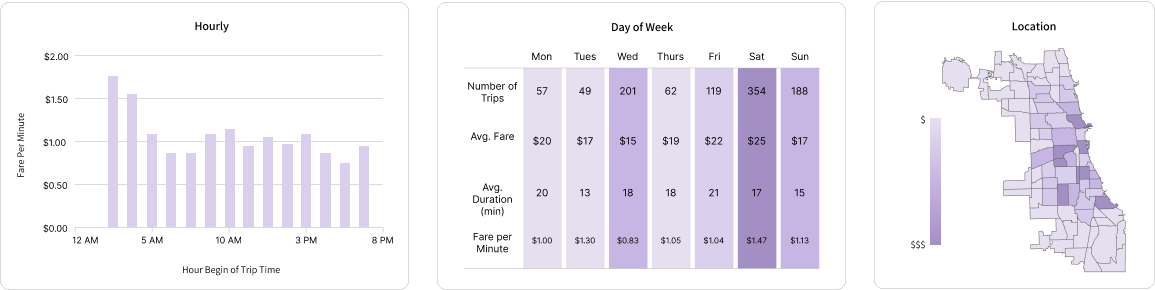}
  \caption{City-level Data Probes}
  \label{fig:aggregatefig}
\end{figure}

\textbf{City-Level Hourly, Calendar, and Map Data Probes.} City-level data probes were all presented side-by-side with their personal counterparts. These were virtually identical to their personal counterparts, only differing in that these versions utilized city-level driver data (see Figure \ref{fig:aggregatefig}). \anedit{For the calendar, weekly calendars were used to compare individual and city-level data because the Chicago dataset does not identify unique drivers.} Drivers saw a matrix of 4 variables per day of the week: total number of trips (for the entire city's dataset), average fare, average trip duration (with passengers in the vehicle), and average fare per minute. The map data probe displayed an interactive map of Chicago, color-coded to indicate neighborhoods with the highest average fares per minute.

\subsubsection{Work Planner}
\label{scheduler-description}
\begin{figure*}
  \includegraphics[width=.9\linewidth]{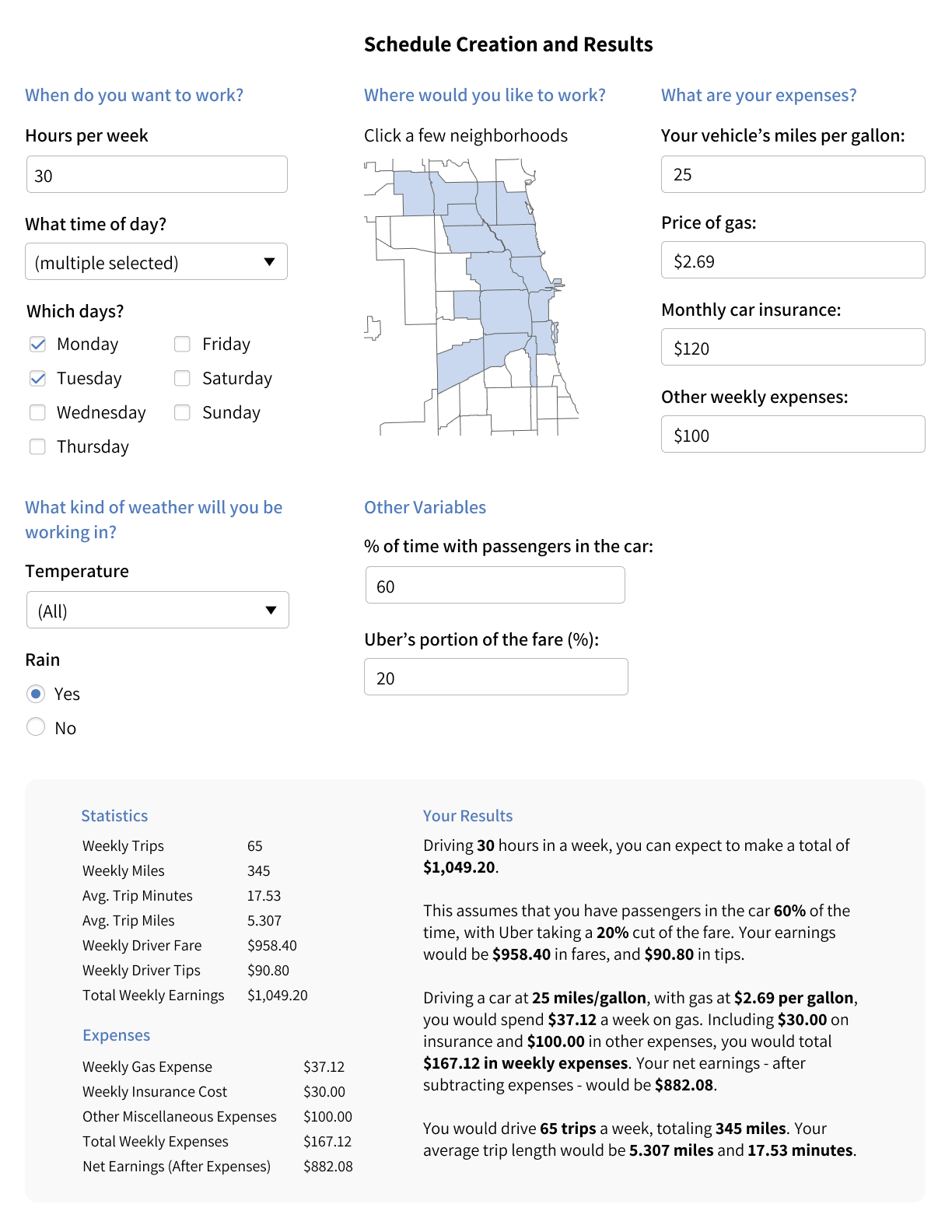}
  \captionsetup{width=0.8\textwidth}
  \caption{Work Planner data probe that participants interacted with to view predictions of their schedules and surface design considerations.}
  \label{fig:scheduler-full}
\end{figure*}

We created the Work Planner (Planner), a comprehensive simulation of predicted earnings and expenses using the aforementioned Chicago data. \anedit{Figure \ref{fig:scheduler-full} displays the full tool details including input and output variables.} 

To use it, drivers input driving preferences: the number of hours driven in a hypothetical week, days of the week driven, hours of the day driving, neighborhoods they choose to pick up in, outdoor temperature, and precipitation. \anedit{Weather variables were included based on past drivers sharing they believe weather may impact their earnings.} Drivers also enter expenses including the price of gas, their vehicle's mileage, their car insurance cost, and other miscellaneous expenses. \anedit{Expense calculation, something rideshare platforms currently lack, was included to give drivers a more realistic sense of their net profit.} They also adjusted other variables important to the simulation including \aledit{the} portion of their fare that the platform withdraws, and the percent of their time spent with passengers in their car. \anedit{The first was included to respect drivers' personal beliefs of how much platforms take which often differs from what platforms formally claim. The second was included as drivers previously warned their hourly earnings rate includes unpaid time spent looking for passengers or driving to them.} For these two variables, we set the default parameters to 25\% and 55\%, standard values for Chicago \cite{manzo_bruno_2021}.

\anedit{Output statistics are generated by using the participant inputs to filter the city-level data to the subset of trips which match these. For example, to calculate weekly earnings, the Planner first calculates the average fare ($AF$, \$) and average trip duration ($ATD$, minutes) directly from the data subset. Next, it determines the number of projected trips ($PT$) through the equation: 
\begin{equation}
    PT = \frac{60}{ATD}*TPC*HPW
\end{equation}
where $TPC$ refers to the percentage of time a driver spends with passengers in the car and $HPW$ represents the hours per week the driver inputs as working. Finally, the weekly earnings ($WE$, \$) is obtained: 
\begin{equation}
    WE = AF * PT
\end{equation}
 Similar calculations are done to obtain the other statistics.

We explained to participants that the calculations were based on June 2022 data from only the Chicago area. We also explained that based on their selections for time and location of work, the tool’s output would adjust accordingly to take into consideration trips started on those days, hours, or neighborhoods.}

\begin{figure*}
\includegraphics[width=.79\linewidth]{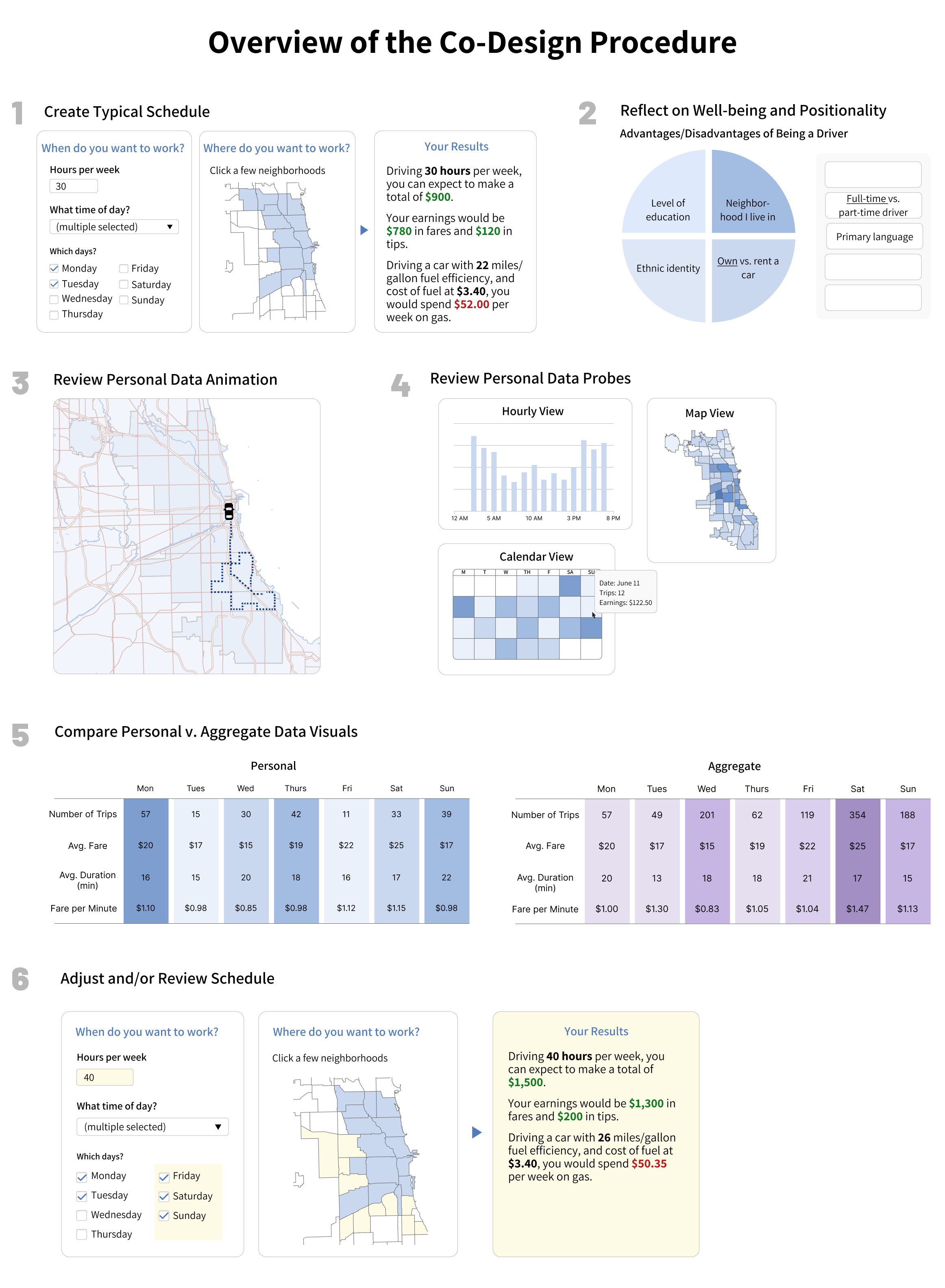}

  \caption{\textbf{Co-Design Procedure:} {\normalfont 1. Participants interact \anedit{with the Planner} to input parameters of their typical work schedule and expense-related information. They review their predicted earnings for whether it aligns with their real experiences. 2. Participants discuss well-being and positionality to identify well-being goals they have and the advantages or disadvantages they face as drivers. 3. Participants view an animation of the trips they drove on a map of Chicago, using their own driving data. 4. Participants review \anedit{individual data probes} to identify trends and/or reflect on what affects their work habits. 5. They compare \anedit{individual data probes} with \anedit{ones} made from Chicago city-level \anedit{driver} data to reflect on similarities and differences with the \anedit{city} driving population. 6. Participants create another work schedule based on the trends they identified from the \anedit{probes} and compare the results with \anedit{the schedule} from step (1) to review which satisfies their preferences.}}
  \label{fig:overviewfig}
\end{figure*}

\subsection{Co-Design Procedure}
Participants began with the Planner to create their typical weekly schedule. Next, we introduced them to the three types of well-being and asked whether they wanted to improve any of them in their work. Participants completed a positionality activity, where they considered 1-4 factors that advantage them and 1-4 factors that disadvantage them in their work from a set of factors often used in positionality exercises \cite{noel2021learning} in addition to ones related to gig work. The section concluded with participants answering, "Is there anything you’d like to know or want to do to improve your work strategy?"

Participants then explored data probes created using their personal Uber data. The first probe was an animation of their Uber trips \anedit{that they watched} as they described their typical driving strategy. The second probe was their hourly bar chart of the earnings they made during June 2022\footnote{For P4 and P10, this was modified to their \aledit{entire} 2021-2022 data because of the sparsity of their trips as part-time drivers}. The third probe was the calendar, from June to their most recent available data, where days were shaded based on their total earnings. Finally, we showed participants \anedit{their map data probe} reflecting their pick-up and drop-off locations and fare patterns \anedit{for} each neighborhood based on the past 30 days. 

With each probe, we guided participants through task-based questions (e.g., "Which days does the calendar show you made the most money?"), followed by questions to understand their patterns, the factors that influence them, and the impact on their well-being. Participants could also hover over each probe to review metrics about their work (see supplementary information \anedit{to view full metrics per probe} in the hover). 

Next, participants compared their individual data probes with corresponding ones created from Chicago drivers: \anedit{hourly, calendar, and map data probes.} Participants viewed their individual and city-level hourly data probes side-by-side and were asked to identify differences and similarities, and reasons for why they believed those exist. This continued for the \anedit{calendar} and \anedit{map} data probes.

Finally, participants created a new schedule using the Planner based on the insights they drew from the data probes about optimal hours, days, and locations to work. They compared the results of their original and new schedule and shared their reflections and preferences between the two. Figure \ref{fig:overviewfig} shows the co-design session flow. The session concluded with participants reflecting on their experience with the five data probes, thinking specifically about how the probes helped them and which visualizations they liked the most.

\subsection{Analysis}
All co-design sessions were conducted and recorded using Zoom, and transcribed using Otter.ai. Each session lasted two hours on average, and participants were compensated with a \$90 gift card. Notes and transcriptions were reviewed by the interviewers and notetakers to produce summaries for each participant. These summaries (containing notes, memos, and reflections) were clustered based on protocol areas and data probes. The data was then analyzed following \citet{patton2014qualitative}'s qualitative data analysis method. Two researchers coded the data and grouped them for emerging themes. The entire research team then discussed these themes to derive the final themes reported in this paper.
\subsection{Background on Uber Features}

We provide a brief explanation of Uber platform features referenced by our participants to provide clarity for our findings. \textbf{Quest promotions} are Uber's incentive-based method of encouraging drivers to increase their short-term driving\footnote{https://help.uber.com/driving-and-delivering/article/how-does-quest-work?nodeId=3a43fa72-4fc2-42d0-bc1d-63c4c0bddb9d}. Quests offer drivers a \anedit{limited time} incentive for completing a set number of trips; e.g., \textit{'Complete 60 trips between Friday and Sunday for a \$150 bonus'}. Quests appear unevenly, with varying thresholds and rewards. Many drivers suspect that Quests unfairly favor newer or inactive drivers. However, because of the opacity of the promotions, they lack the means to definitively prove this suspicion \cite{zhang2022algorithmic}. \textbf{Surge promotions} are Uber's method of matching the supply of drivers with the demand from riders\footnote{https://www.uber.com/blog/courier-surge-intro/}. Uber will temporarily create "surges" in high demand areas to incentivize drivers to relocate to those areas. \textbf{Uber Pro} is Uber's driver rewards programs that automatically enrolls all Uber drivers\footnote{https://www.uber.com/us/en/drive/uber-pro/}. There are four tiers of Uber Pro: blue, gold, platinum, and diamond. Driver tier is determined by a variety of factors including driver rating (the average of their last 500 ratings from passengers), cancellation rate, trip acceptance rate, and the number of 'points' earned. Points are re-calculated every 3 months, and earned by completing trips with certain high-demand times earning more points per trip. 
\section{Findings}
\anedit{In co-design sessions, we observed that data probes acted as boundary objects for participants to effectively communicate their work patterns and contexts with our research team and suggest AI design implications rooted in these contexts. In this section, we describe three main insights surfaced by participants through data probes and corresponding (AI) design considerations. First, using data probes, participants shared how well-being trade-offs and immovable positionality factors pose constraints for how they could use the Planner---the Tableau schedule and earnings simulator we created---practically, but suggested alternative ways for an algorithmic work planning tool to center well-being within their constraints. Second, participants explored data probes to identify factors beyond those represented in their data probes that impact their work and well-being, offering new prediction ideas for AI tools that can support them. Third, participants used data probes to identify specific instances of unfair algorithmic management and suggested ways to repurpose data probes and design algorithmic tools to audit platforms.}

\subsection{Well-Being Trade-Offs and Positionality Constraints That Impact Worker's Strategies}
\anedit{Participants used their data probes to explain how these visualizations reflect the ways they currently balance well-being in their strategy as well as how personal obligations constrain their work flexibility. It is important for worker-centered AI design to be considerate of these limitations in order to provide well-being support that is actually feasible for workers to take.}
\subsubsection{Surfacing Well-Being and Positionality Through Hourly, Calendar, and Map Data Probes}
\hfill \break
\anedit{
\textbf{Initial reflections around well-being and positionality.} When we first introduced well-being and positionality, a handful of participants raised well-being concerns such as safety or exhaustion (P3, P8, P10), stress over precarity of earnings and passenger attitudes (P3, P12), and frustration over traffic or unfair treatment by platforms (P8, P11). For positionality, participants shared advantages that directly impact their earnings such as “Being a Multi-Platform Worker” or reduce their expenses such as “Owning My Car”, and disadvantages that limit their earning ability such as living far away from centrally located, high earning neighborhoods.
\newline

\textbf{Articulating well-being trade-offs and positionality constraints using hourly and map data probes.}
Through data probes, participants surfaced or expanded on initially identified well-beings and positionalities. The hourly and map data probes were especially effective as tools for participants to discuss specific well-being trade-offs they make---such as balancing physical and mental health with financial gain---and positionality constraints they face---particularly familial obligations. These probes were likely effective given participants held safety concerns about Chicago’s high rate of carjackings and shootings (P1-3, P5, P8-12), which are highly associated with late night/early morning hours and specific neighborhoods \cite{cbs_news_2022, city_of_chicago}. Participants explicitly identified high earning hours and neighborhoods on hourly and map data probes of city-level data, saying the probes confirm what they already “know” are profitable strategies. Then, they explained their individual data probes reflect their conscious decisions to not drive at these times or areas because of well-being concerns and positionality limitations. P2 said, "maybe it [working late night hours] could have a positive impact on my finances, rates, more earnings, but I just wouldn't feel safe. And I would feel, you know, anxiety”, and P10 referenced how his fear of his daughter losing her father prevents him from driving in certain areas. In contrast, P4 did not hold the same concerns over his well-being, instead reflecting a high level of empathy towards the positionalities of residents of these neighborhoods. He accepted rides regardless of pick-up or drop-off locations because he did not want to see people lose access to rides simply because they lived in more "dangerous" neighborhoods. 
\newline

\textbf{Surfacing obligations imposed by workers' positionalities using the calendar data probe.}
The calendar data probe surfaced more unique perspectives around the kinds of limitations worker positionalities impose on their work patterns. As people often use calendars to record and view their daily tasks, this probe was intended to help people intuitively recall their data in the context of their responsibilities and routines. Accordingly, while exploring his calendar, P2 explained his data probe showed few working Saturdays and Sundays because he aligns his days off to be with his girlfriend. Comparing the city-level and his individual weekly calendars, P3 lamented being unable to capitalize on high-earning weekends due to caring for his infant child. P1 explored his individual monthly calendar for memorable low- or high-earning days, and expressed regret when he remembered that he had been unable to work on July 4th because of family obligations. P5, P7, P9-10, and P12 also expressed similar constraints in their weekly schedules due to commitments with family and friends. 
}

\subsubsection{Design Consideration for Centering Well-Being in AI Design}

\hfill \break
\anedit{
\textbf{Developing assistive AI for tracking and reflecting on well-being goals.} While exploring their data, several participants commented how they try to prioritize their physical and psychological well-being with days off but are not always successful. For example, P5 told us Sundays are family days, yet his calendar showed him working half of all Sundays. P6 tries to prioritize time with family and friends as he gets older, but rather than treating it as time off, he feels he must make up for it later. P2 expressed surprise when he saw his calendar, explaining he thought he only worked 15 days in June when in reality, he worked 25 days in the month. Data probes illuminated the difference between what drivers believe they do for their well-being versus what occurs in practice. AI tools can focus on supporting drivers in the same regard. This could be through an AI work planning tool which uses a driver’s data to present statistics about recent days worked so they can see whether/how they are meeting well-being goals and how to integrate that knowledge into the next week’s plan. 
}

\anedit{AI tools could also assist drivers with well-being centered schedules by predicting when to take breaks or what past factors led to profitable days. Unsurprisingly, long-tenured drivers felt confident in their driving strategy and did not seem receptive to change. But a few mentioned an interest in using their data to influence their strategy. P6 wanted to use his data to predict profitability so he could schedule appointments or breaks on days with lower predicted earnings. P12 shared how he would like a predictive model to help him decipher what factors led to him having a good or bad day (in earnings), "As a driver, you can analyze it and be like, 'Man, I've really had a bad day. Why did I have such a bad day?' Or 'I had a great day. What is it that made it such a great day?'" Similar to P6, P8 was interested in using her data to plan her day off, adding the benefit of taking time for herself: “I tend to do better with tips back after a couple days off, which means I'm just getting burned out of people.” Addressing her comment, AI tools could also use workers’ own data to enable them to track and reflect on their time off and see when they may benefit from a break.
}
\anedit{
\newline

\textbf{Respecting driver contexts and knowledge.} For most participants, viewing data probes did not lead to changes in their work strategies due to immutable constraints---not driving at night due to safety concerns despite the potential profit or being a caretaker limiting the time they may work. However, they gave suggestions for how to design AI tools that can assist them in decision-making for well-being. This is a reminder that a tool designed without considering worker contexts may very well violate their personal constraints or expertise and will unlikely be adopted. Instead, AI tools should take on assistive roles, leaving final decision-making to workers themselves. For example, by their own admission, drivers get burned out, thus a tool that can track and help them reflect on the effect of taking time off could be more effective in advancing worker well-being than one which prescribes to them to take a break.
}

\anedit{Additionally, we observed that for a few participants, the session did lead to recognizing potential changes to improve their well-being. P2, though an experienced driver, described himself as being in a “transition period”: he was moving and his girlfriend recently changed her days off, so he was developing a new schedule. Though confident in the merits of his old schedule, he showed interest in using the Planner to simulate how to tweak his schedule. P1 and P11 both considered testing new schedules which estimated fewer earnings but also meant fewer or different hours in order to improve their psychological and physical well-being. Though they were not interested in change, P8 and P12 shared that reviewing the data probes and using the Planner could help beginner drivers instead of established drivers. Once again, through understanding their context, we can more appropriately determine the use and audience of tools like algorithmic work planning. Rather than catering to general users, a work planning tool could be designed to center around onboarding new drivers or experienced drivers with an existing interest in tweaking their strategy.
}

\subsection{Identifying New Factors Not Included in Data Probe Design}
\anedit{Through data probe exploration, participants identified additional factors which impact well-being as well as ideas of predictions they wanted to see from AI tools. Participants also used the data probes to recognize and explain gaps in their Uber data which resulted from other platform work.}

\subsubsection{Identifying Additional Factors that Impact Well-Being Through All Data Probes}
\hfill \break
\anedit{
Our data probes centered around participants reflecting on schedule-related factors of work (e.g., time, day, location). Encouragingly, exploring their data in these ways also allowed participants to remember factors we did not include in the data probes but that they felt were influential on well-being, including passenger characteristics, general work precarity, perceived success of driving patterns, and multi-platform work.}
\newline

\textbf{Passenger characteristics.} Reviewing their hourly data probes, P7 and P9 felt they currently work optimal hours which do not lead to exhaustion or stress. Instead, P7 reflected that far more impactful on his well-being are interactions with passengers---negative interactions can pull his mood down: "No one wants to interact with someone who...is mistreating them or being obnoxious." Additionally, positive interactions with passengers can boost drivers’ moods. By inspecting her calendar data probe for high earnings days, P8 recalled that Saturday afternoons are often profitable due to tourists, adding that a good day is when she can connect with her passengers: “I feel at times like I’m an advocate for my city”. 
\newline

\textbf{Work precarity.} Because of how it allowed participants to view trends in their earnings across days, the calendar data probe led to several participants raising work precarity as another feature that negatively impacts their well-being (P3, P5-6). For them, work precarity led to continuous stress trying to meet financial goals and having to make up the hours elsewhere. They described that this worry often manifests as feeling unmotivated at the beginning of the week and then anxiously trying to make up for it at the end with nonstop work and late hours. Participants illustrated this situation by pointing to the color gradients on the calendars where Mondays and Tuesdays were light in color (lower earning) and Fridays and Saturdays were darker (higher earning). In contrast, part-time drivers P4 and P10 described being unaffected by gig work precarity because the income is not essential so they have freedom to stop and start at their convenience.
\newline

\textbf{Perceived success of driving patterns.} Some associated well-being with how successfully they felt they followed driving preferences or plans. They expressed how not being able to follow these patterns impacted their day. Using the map data probe, P1 mentioned the stress he feels driving downtown versus areas that he likes, such as the airport. P3 explained if he experienced a trip he considered a "mistake", such as driving in the suburbs during rush hour, it could wreck his day. As long as she was able to maximize her efficiency through higher earnings and shorter trip durations, P8 felt comfortable both financially and physically.
\newline

\textbf{Missing data from other work platforms.} Across all data probes, we observed the challenges multi-platform workers (P1, P4, P6-12) face in truly contextualizing their work patterns and understanding their well-being. When reviewing his calendar, P7 explained that he held little stock in the patterns displayed as they were partial information. But if he had data from all his platforms, he could make “a better determination of what I'm doing, and what I'm making and why that particular time and particular day.” Still, we found that even in the presence of missing data, reviewing their probes to identify hours or days they had very little to no earnings helped many recall switching to another platform due to better bonuses or tasks, and expand on their typical patterns for multi-app work (P4, P6-P12). 

\subsubsection{Design Consideration For Algorithmic Work Planners or Other AI Tools.}

\hfill \break
\anedit{

\textbf{New predictions to make to address well-being factors.} Some participants suggested new things they wanted to see predicted in the Planner. P11 suggested predicting passenger tips which directly impacts their total income, echoing the comments by other participants around how the uncertainty of passenger behaviors greatly affected their moods (P7, P9). P11 also suggested predicting optimal starting times and places to support drivers in structuring their work. This could also help address the challenges drivers face with work precarity by anchoring them with a starting point to stay on track each day. P5 also made a suggestion for how to help workers stay on track with schedules, explaining how he was interested in having upcoming purchases or expenses integrated with a calendar data probe. Based on his comment, an AI tool could assist drivers in simulating more control around their work plans by taking their historical data and future spending to provide feedback for how long it may take to meet purchase goals.
\newline

\textbf{Using data probes to understand gaps imposed by missing platform data.} Ideally, tools to support workers will integrate data from all platforms they work to give a comprehensive understanding and provide useful support for future decision-making. However, this is not always possible because of platforms’ different policies for providing data to workers. In lieu of this, we observed how data probes can play a role to at least fill some gaps of understanding left behind by missing platform data. We recall how P8 used her hourly data probe to review strange patterns and recall specific multi-platform behavior: she had no Uber data for June 2022 on specific hours (e.g., 1pm, 3pm) and remembered taking advantage of a bonus Lyft ran on odd afternoon hours that month. One common problem for data practitioners is determining how to handle missing data when building models---working with workers can surface specific reasons for why missing data exists (e.g., promotions on another platform), and how to handle it as informed by the workers’ own experiences. 
}

\subsection{Instances of Unfair Algorithmic Management}
\anedit{
Algorithmic management came up frequently and unprompted during sessions as participants discussed how they lack control over their strategy in the face of unfair platform algorithms. Using data probes, participants identified instances of algorithmic management and imagined how probes or other algorithmic tools could help them resist or investigate algorithmic management. 
}

\subsubsection{Demonstrating Unfair Algorithmic Management Through Animation, Hourly, and Map Data Probes}
\hfill \break
\anedit{
Although some participants used the Planner and found their new schedule promising to explore further, others (P3, P5, P11-12) lamented that the changes they made for locations to drive in are impossible because Uber does not allow them to entirely control the locations of trip drop-offs they are offered. This lack of control is related to heavy-handed algorithmic management identified by drivers in prior work \cite{mohlmann2017hands,zhang2022algorithmic}.
\newline

\textbf{Identifying unfair algorithmic management with the animation and map data probes.} The animation and map data probes were integral to participants identifying trips that deviate from their usual preferences. Though P3, who started less than a year ago, characterized these deviations as made "probably by accident", others discussed these trips as a result of manipulative algorithmic management (P2, P5-P6, P10-11). For them, their probes demonstrated how platforms reduce driver autonomy by assigning them rides that go against their preferences and well-being. Watching his animation, P10 expressed frustration over how platforms seemingly intentionally send him rides in the opposite direction of where he wants to go at inopportune times: "Yeah, I remember that ride…I wanted to end the night and get home, and then of course, it's like that always seems to happen where it's like Uber or Lyft just happen to know that...So then that was a struggle to get back home to find rides, you know, at 2, 3 in the morning to get back. It's not easy." 

Some participants mentioned that Uber does allow drivers to set two destination filters a day to provide them \aledit{with} a limited level of control. However, quite a few shared that these destination filters “don’t work” or have stopped working in the past month (P3, P8, P10-11). P9 mentioned the possibility of Uber treating drivers differently depending on status, saying he believes the trip assignment algorithm favors him for his high Uber status (i.e., little downtime between ride requests) but that a driver of lower status probably experiences high downtime between requests. If true, this hypothesis could also apply to the disparate experiences of the destination filter amongst our participants.
\newline

\textbf{Identifying platform manipulation with the hourly and map data probes.} A couple participants mentioned how the data probes reflect Uber’s manipulation tactic to keep people driving on the platform as long as possible. When asked what influenced his driving patterns as shown by his hourly data probe, P3 explained them as the product of Uber’s Quests. Whereas he would otherwise sign out during slow hours to take a break, Uber pressures him to keep driving in the attempt to complete Quests. Reviewing the map data probe, P6 talked about how he tries to stay close to places where surges often occur. He explained that from past mistakes, he has learned not to chase surges after realizing it is a bait and switch tactic Uber uses to lure drivers to a certain region and then retract the surge bonus.
}
\subsubsection{Design Consideration for Algorithmic Worker Tools To Combat Algorithmic Management}

\hfill \break
\anedit{

\textbf{Validating hunches of algorithmic unfairness through quantitative data.} Viewing the data probes, participants gave suggestions that point to how tools or data probes can support drivers against algorithmic management. Participants liked generally being able to see their strategy and movement patterns on their animation in a visceral manner. P1-2 and P6 added how it helps them quantify their hunches about their movement patterns and where trips take them. P1 and P6 specifically shared that the animations confirmed their beliefs that platforms regularly send them rides to places they do not want to go: "So that's kind of a nice way to just see, oh, I do end up---this isn't my imagination. I am getting sent on these trips that take me here all the time." (P6). These suggestions for using data probes to provide evidence points to an excellent direction for tools to support drivers: How can we use participant data so drivers can investigate their hunches of platform manipulation in a measurable way?
\newline

\textbf{Reverse engineering platform algorithms.} P6 also shared an idea around how to help drivers predict surges through reverse engineering Uber’s surge algorithm. He observed that the passenger Uber app seems to incorporate surge pricing a few minutes before the drivers see it, thus he said a tool to help drivers effectively predict surge would need to connect to the “rider side of the app that’s showing where...the demands are typically highest”. He explained that some drivers already track this information through notebooks and screenshots, although he does not personally track it. 

These ideas of reverse engineering and platform auditing through data echo work done by \cite{calacci2022bargaining} where workers may be able to collectively share and analyze data to identify widely-encompassing platform manipulation or unfairness.
}
\section{Discussion}

Our findings suggest that participants were able to use the different data probes to help them identify and communicate work patterns as well as personal or situational contexts that impact their strategies. We discuss the implications \anedit{on the design of data probes as well as using} data probes \anedit{to} design AI with stakeholders and for worker advocacy.

\subsection{How to Use and Improve Data Probes}

\subsubsection{Reflections From Co-Design Sessions with Data Probes}
\anedit{
One of the greatest benefits of using data probes is the ability to surface situational contexts or limitations workers face which may not be obvious from their data alone, similar to \cite{bogers2016connected}'s use of diary entries and interviews to understand raw sensor data. Used without context, the data in an AI tool could lead to inappropriate recommendations. For example, the data alone could identify what drivers told us is a trend: the slow hours for work are typically 10AM-2PM. However, for P1, these hours overlap with when he stops working to begin caretaker duties for his father. If a tool were created to suggest the best hours for P1 to work, it might assume he chooses not to work these hours because of low demand and fail to treat it as a hard constraint for him. 

Using multiple data probes was also crucial, not just to prompt different insights, but because it often took repeated mentions of well-being or positionality before participants fully revealed their circumstances. This may be due to a need to establish trust during the session. For example, P1 did not select caregiver as a positionality trait, nor did he mention it during the hourly probe though it restricts his hours. Yet he disclosed this limitation during the calendar probe. Using different probes may also be necessary to dislodge worker notions of what valid well-being concerns are. P9 initially described his well-being as “physically, I’m really good, financially, I’m making good money, and psychologically, I’m perfect”. But discussing how he feels driving late night hours, it appeared that his view on well-being is influenced by his concern of platform punishment: he explained he strategically masks physical exhaustion from passengers out of fear of platform deactivation. 
 }
\subsubsection{Improvements for Data Probes}
\anedit{
In order to better compare themselves with Chicago drivers, participants suggested different metrics to display and subsets of driver data to view. Though we displayed metrics such as average fare or miles per trip or per minute, participants associate different metrics with different types of strategies: some people try to maximize earnings per trip while others try to maximize total number of trips. P2 follows the latter and wanted to compare number of drop-offs with all Chicago drivers, explaining that “you can have really high average trip per min average, but maybe not as high as in terms of the number of trips.” Participants also suggested comparing themselves to similar subsets of drivers rather than all Chicago drivers including experienced full-time drivers only (P6, P8), type of car (P8), and Uber Pro status (P9). P4, based in a suburb, and P8, based in the city, suggested comparisons filtered by trips originating from the suburbs or city would be more reliable because of inherent differences: city trips are lower-earning but fast, while suburban trips are higher-earning but longer. P4 remarked, "While I'm lumped in Chicago data, I'm not Chicago. It's a completely different animal".

Participants also gave suggestions to improve how features are defined in data probes. P2 and P9 reminded us the Planner should consider car expenses that occur at non-weekly intervals such as car maintenance. Nearly all participants said our current day and hour inputs for the Planner are too limiting as these features vary depending on the day and week. Finally, exploring the calendar probes led participants to point out the need to align how data probes define begin/end of days and weekdays versus weekends with how Uber, and consequently drivers, define them (P1, P4, P9-10)---e.g., drivers define weekends based on weekend Quests which usually run Friday 4AM-Monday 4AM.

We propose that these improvements for data probes also translate to AI design considerations due to similarities in characteristics they present---e.g., suggestions for redefining feature boundaries align with feature engineering. Also, while some features like expenses may be knowable without data probes, others may not necessarily surface naturally without tools like data probes. For instance, in an interview, it may not seem obvious to ask drivers (or anyone), “How do you define the hours of a day/weekend?” Misunderstandings might be perpetuated if data analysis tools are built using flawed logic (which is not normally displayed to users). Instead, in our sessions, participants explored data probes to view hourly breakdown on the days in the calendar, allowing them to correct us on how to define a deceptively simple feature.
}

\subsection{Using Data Probes to Elevate Stakeholder Expertise in AI Co-Design}

First, we share the realization we made in our findings around our own assumptions about drivers in order to demonstrate the importance of building tools \textit{with} and not \textit{for} them. Specifically, we discuss how positionality and well-being shape work preferences and constraints participants hold, summarizing our findings around the ideas and contexts they surfaced while exploring their data probes and how these can inform AI design.

Participants' personal contexts and situational factors we learned can inform directions and challenges for AI practitioners and UX designers to consider. One of the implications for AI design from our co-design sessions is the importance of surfacing hard constraints for AI products that support workers. As we learned in our sessions, participants’ positionalities did not always reveal themselves during our well-being and positionality activities, but became more explicit \textit{through} the exploration of data probes. These hard constraints are also informative for AI designers in the consideration of tools---for example, in the design of nudges, to show recommendations that respect what a stakeholder is able to do within their confines, designers must surface and integrate these nuanced hard constraints.

Prior work has explored different approaches for modelling and forecasting passenger travel behavior to predict where passenger demand will occur within certain time intervals \cite{chen2021short, niu2018real}, even using external features such as weather, demographic data, and crime rates \cite{chen2021short}. However, these studies focus on demand as the principal element of passenger behavior. Future work can be done to predict other passenger characteristics such as tipping habits, travel patterns, and propensity for conversation depending on the availability of passenger data and issues of data privacy and confidentiality. Of course, when devising and integrating these insights, care must also be taken to ensure that they do not become a means of discrimination against certain groups of passengers and exclude them from access to rides as a result.

Treating workers as experts can also help AI practitioners and designers recognize challenges to consider that they may not have thought of before. For example, participants' inconsistent work patterns and concerns around Uber's constant app changes would impact the \anedit{patterns exhibited in their} data used for AI tools. We recall how P8 and P11 have to alter their work strategy often due to Uber's frequent changes. We also recall how multi-platform drivers switched working between apps in inconsistent ways and had gaps in their Uber data due to working on multiple platforms. These problems that participants identified are reminiscent of the data drift problem, where there is a mismatch between the data that a model was trained on and future data. To help augment those explorations in this domain, workers can \anedit{provide additional explanations to} contextualize how or why their patterns are changing, based on the platform app changes or their own personal situations. 

Participants' suggestions of additional factors for the Planner or new predictions the Planner could provide \anedit{as discussed before further support the value of AI co-design with stakeholders.} Based on their everyday driving, participants were able to share what has influenced their driving patterns and earnings in the past, including \anedit{a driver's neighborhood and commuting tendencies}, large events that lead to high surges, and weather or other seasonal data. Participants also shared ideas for what would be more useful as predictions for them than the earnings we displayed, such as optimal start location and time, and tipping probability of passengers in different neighborhoods. These participants' ideas \anedit{reiterate} the value of incorporating them as experts in AI design to promote the co-design of worker-centered tools.

\subsection{Rideshare Worker Advocacy}

Based on our findings, we propose there is potential \aledit{for} the use of data probes as boundary objects to work with researchers, advocates, and policymakers advancing worker rights. Collective driver goals typically consist of increased algorithmic and data transparency \cite{siddiqui2019uber, khovanskaya2019tools}, greater and guaranteed pay\footnote{https://www.drivers-united.org/}\cite{lazzaro2016uber}, and increased access to benefits such as healthcare, wellness support, and education reimbursement\footnote{https://il.driversguild.org/about-us/}.

We observed how the data probes were very effective tools for workers to communicate complex ideas and trends with us, and for us to understand their work strategies and how they are shaped. Several participants (P1, P6, P10) noted during their session how their data probes allowed them to quantify or prove beliefs regarding algorithmic management. The animation allowed them to quantify when Uber was deliberately not honoring the in-app destination filter feature that Uber claims helps drivers target locations they want to go to. The combination of how the data probes worked as boundary objects for us with participants and how the data probes helped participants identify instances of manipulative algorithmic management leads us to propose the data probes we have developed as methods that other researchers, advocates, and policymakers working with gig workers can adopt to advance worker rights. 

\subsubsection{Boundary Objects for Advocacy}
The tools designed for this study could serve as a framework for future tools designed for non-driver audiences. Policymakers have a limited understanding of platform practices, and because of Uber and Lyft's opaque processes and limited data availability, it can be difficult for governments and regulatory agencies to get an accurate grasp of the real-world driver context. However, local governments and courts are increasingly issuing legislation and decisions requiring rideshare companies to publicize their data, including New York in 2017 \cite{morris2017new}, Seattle in 2018 \cite{brasuell_minkoff_2018}, and San Francisco in 2019 \cite{said2019uber}. Because of the increased availability of data, it is becoming more feasible to build similar visualizations for different cities and at a larger scale. This type of interactive tool allows a depth of understanding beyond the typical statistics seen in a white paper or fact sheet, acting as a boundary object between drivers and policymakers. This can advance drivers' collective interests by adding more accessible public-facing data-driven tools to raise public awareness of driver issues.

While some cities and states have established minimum wages for drivers such as New York City in 2018 \cite{campbell2018new}, California in 2020 \cite{lyons_2020}, and Washington in 2022 \cite{bellon_2022}, the lack of legislation in other locations---such as Chicago---means that drivers are often being paid less than local minimum wages after subtracting expenses \cite{roeder_2022, manzo_bruno_2021}. Countless studies have been performed examining rideshare datasets \cite{manzo_bruno_2021, parrott2018earnings}, however, without publicly available data, this type of independent analysis is impossible. However, there are still problems with data-driven studies as a means of activism. Two prior studies attempted to calculate average hourly earnings in Seattle and came up with two very different values (\$23.25/hour vs. \$9.73) due to wildly differing methodologies in calculating expenses \cite{marshall_2020}. Intense care must be taken regarding understanding the methodology and meanings of ridesharing terms before issuing declarative statements or legislation on the basis of data. Through our partnership with the Independent Drivers Guild, we hope to develop further public-facing tools to build awareness of the realities of rideshare work and advance worker rights.

\section{Limitations}
We acknowledge that our study has limitations. We faced challenges in recruiting a diverse range of drivers, possibly due to the requirement in our study for participant data; our participants came primarily from Reddit, Facebook driver forums, or a driver advocacy group, thus they may be more data literate and aware of platform issues than a standard driver. We conducted this study with Chicago drivers where we had access to publicly available data. Additionally, due to the limitations of Uber's data, we were unable to calculate certain metrics for drivers. The duration of our sessions limited how much participants could share with us. Our sessions lasted for two hours, yet some participants wanted to discuss more after it ended. Finally, as a qualitative study, the findings we offer around worker-centered AI design should be further investigated through other research methods with a larger number of participants. 

\section{Conclusion}
Using workers' own data, we designed data probes and conducted co-design sessions with workers to surface their work patterns and how those are affected by their well-being and positionality. \anedit{We discovered these data probes functioned as boundary objects to help participants surface three main insights: 1) participants described well-being trade-offs and positionality factors they face in their work, 2) they identified factors not in their data probes which impact their work and well-being such as unpredictability of passengers, and 3) they used data probes to identify instances of algorithmic management. For each finding, we also share the design considerations raised by participants using their data probes.} We discuss the implications of \anedit{our study on the design of} data probes \anedit{and the use of data probes to} elevate worker expertise in AI design, including revealing individual contexts to inform the constraints of AI products, as well as the potential for data probes to be used in the future to support worker advocacy efforts.
\begin{acks}
This research was partially supported by the following: the National Science Foundation CNS-1952085, IIS-1939606, DGE-2125858 grants; Good Systems, a UT Austin Grand Challenge for developing responsible AI technologies\footnote{https://goodsystems.utexas.edu}; and UT Austin’s School of Information. We are grateful to Mark Smithivas for his indispensable counsel that guided our research direction, Bianca Talabis for creating the illustrations in this paper, Charles Miele and Yihan Xi for their assistance in creating the animation data probes, the anonymous reviewers who provided invaluable feedback, and our participants for their trust in our work and thoughtful insights during sessions.
\end{acks}

\balance
\bibliographystyle{ACM-Reference-Format}
\bibliography{references.bib}
\onecolumn
\section{Appendix}
\begin{figure}[!ht]
\captionsetup{width=0.7\textwidth}
\includegraphics[width=.5\textwidth]{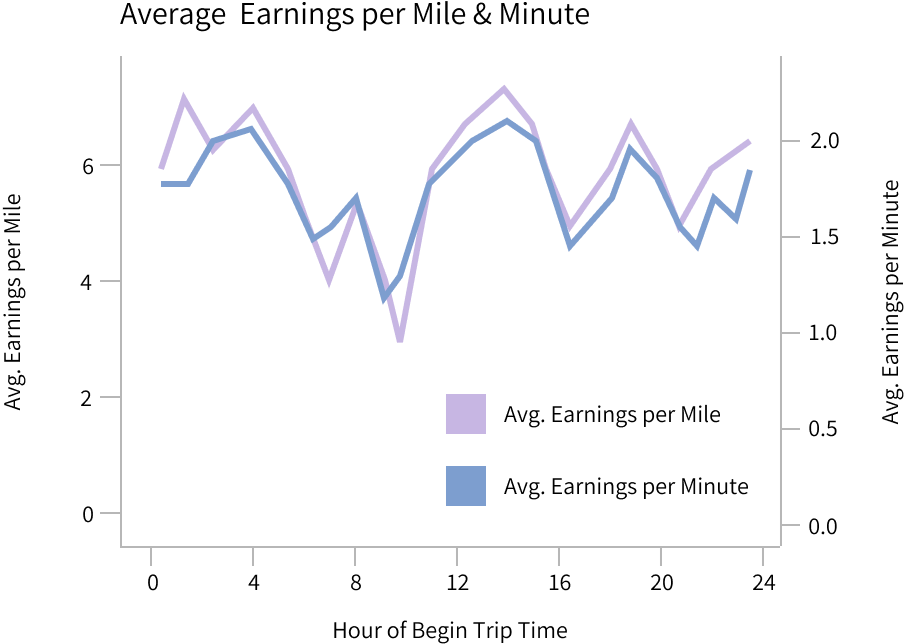}
      \caption{One of the figures created using driver's individual data for the pilots. This uses their trip history to depict average earnings for the time and distance they drove in trips.}
    \label{pilot1}
\end{figure}

\begin{figure}[!ht]

\captionsetup{width=0.7\textwidth}
\includegraphics[width=.8\textwidth]{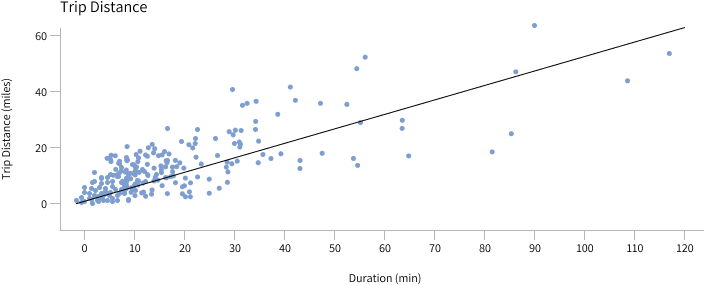}
      \caption{One of the figures created using driver's individual data for the pilots. This uses their trip history to plot each trip for its duration and distance so drivers can identify patterns or outliers of trips they drive.}
    \label{pilot2}
\end{figure}

\end{document}